\documentclass[draft]{aipproc}

\layoutstyle{8x11double}

\begin{document}

\title{Beyond linearized cosmology}



\author{Dirk Puetzfeld}{
  address={Department of Physics and Astronomy, Iowa State University, Ames, IA, 50011, USA}, 
  email={dpuetz@iastate.edu}, 
  homepage={http://www.thp.uni-koeln.de/$\sim$dp}
}

\begin{abstract}
We comment on the necessity of a unified approximative scheme within relativistic cosmology which would allow us to classify different cosmological models in a systematic way. We also report on recent progresses in formulating a cosmological post-Newtonian approximation and the problems related to such a scheme.
\end{abstract}

\maketitle

\section{Motivation}

Approximation schemes have played an important role in the development of General Relativity (GR) since its formulation back in 1915, cf. \cite{Einstein:1916:2,LorentzDroste:1917,EinsteinInfeldHoffman:1938,Fock:1939,EinsteinInfeld:1940,EinsteinInfeld:1949,Papapetrou:1951,Papapetrou:1951:2,Fock:1959} for some early approaches. Several schemes have been devised to quantify relativistic gravitational effects in the context of isolated gravitating systems, thereby permitting a confrontation of the at that time new gravitational theory with experiments. Even many modern applications, which take into account general relativistic effects, rely on approximative methods \cite{Soffel:1989, Brumberg:1991}. 

Nowadays it is common practice to assess the success of new gravity theories by formulating them in a standardized parametrized language \cite{Will:1993}, which allows for a rapid comparison with the outcome of several experiments. Despite the ample use of such a scheme in the context of isolated gravitating systems no such formulation exists in a cosmological context. The lack of such a scheme is rather surprising since, on a first guess, one would expect that cosmological scales demand for a relativistic treatment and, therefore, for an approximation which allows us to classify different cosmological models in a systematic way. 

In the following we outline the program which has to be carried out in order to formulate such a framework in cosmology and report on recent progresses in formulating the first order post-Newtonian (1PN) cosmological hydrodynamic equations for Einstein's theory.

\section{Approximation schemes}

Although there exists a whole host of different approximation schemes in relativistic gravity we are going to concentrate only on two in the following. At this point we also have to remind the reader that there is no commonly accepted definition of these schemes, and, therefore, the usage of the names for them is far from unequivocal. In order to fix our usage of the terms we briefly introduce the so-called {\it post-Minkowskian} (PM) and {\it post-Newtonian} (PN) approximation. 

In the post-Minkowskian or weak-field / fast-motion approximation one extends the gravitational potential, i.e. the metric in Einstein's theory, around the flat Minkowski metric\footnote{Greek indices shall run from $0,\dots,3$ and latin indices from $1,\dots,3$.} $g^{\mu \nu } =\eta ^{\mu \nu }+ \epsilon \gamma ^{\mu \nu }+ \dots$. In many applications one then tries to rewrite the field equations of GR in a form which closely resembles the form of the inhomogeneous wave equation, c.f. \cite{Kerr:1959:1,Kerr:1959:2,Kerr:1959:3,AndersonDecanio:1975,Anderson:1976:2,Ehlers:1977}. The solution of the field equations then being constructed by the use of a retarded Green's function. No other matter variables are expanded in the PM approximation. 

The strategy in the post-Newtonian or weak-field / slow motion approximation is different. Here one develops the metric, starting from the Newtonian limit of Einstein's theory denoted by $\stackrel{0}{g^{\mu \nu}}$, into a series of inverse powers of the speed of light, i.e.\ $g^{\mu \nu } = \stackrel{0}{g^{\mu \nu}}+ c^{-1}\stackrel{1}{g^{\mu \nu}}+\dots$. In addition the four velocity, which enters the definition of the energy-momentum tensor, is developed into a series $u^{\mu} = \stackrel{0}{u^{\mu }}+ c^{-1}\stackrel{1}{u^{\mu}}+\dots$ in inverse powers of $c$. The strategy in solving the field equations in the PN approximation also differs from the one employed in the PM approximation. The goal in the PN approximation is to recover a set of equations which resemble the form of the field equation in Newtonian gravity, i.e.\ Poisson's equation. Since GR has more structure than Newton's theory, one usually has to deal with a coupled set of Poisson like equations in the PN approximation.  

Different ways of labeling the post-Newtonian order, i.e.\ counting the powers of $c$ up to which the metric components need to be determined, can be found in the literature. Here we stick to the counting scheme originally introduced by Chandrasekhar and coworkers in a pioneering series of works on the PN approximation for isolated gravitating systems, cf.\ \cite{Chandrasekhar:1965:1,Chandrasekhar:1965:2,Chandrasekhar:1969,ChandrasekharNutku:1969,ChandrasekharEsposito:1970}.

\begin{figure}
  \includegraphics[width=.6\textwidth]{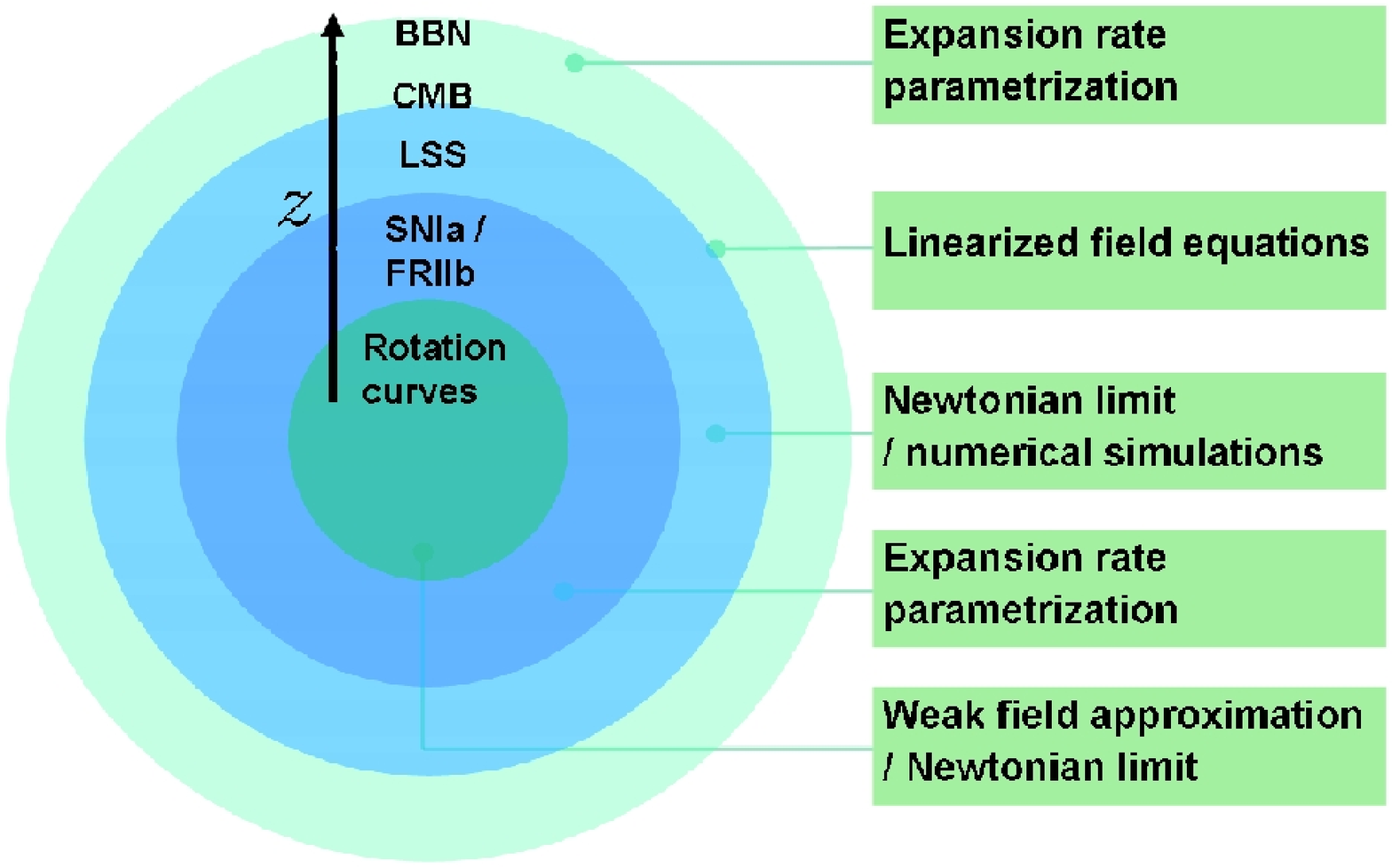}
  \caption{The current cosmological tests make use of different approximation schemes. So far we do not have a unified framework which would allow for a systematic comparison on the different scales currently tested by cosmology. }
  \label{scales_fig}
\end{figure}

\begin{figure}
  \includegraphics[width=.7\textwidth]{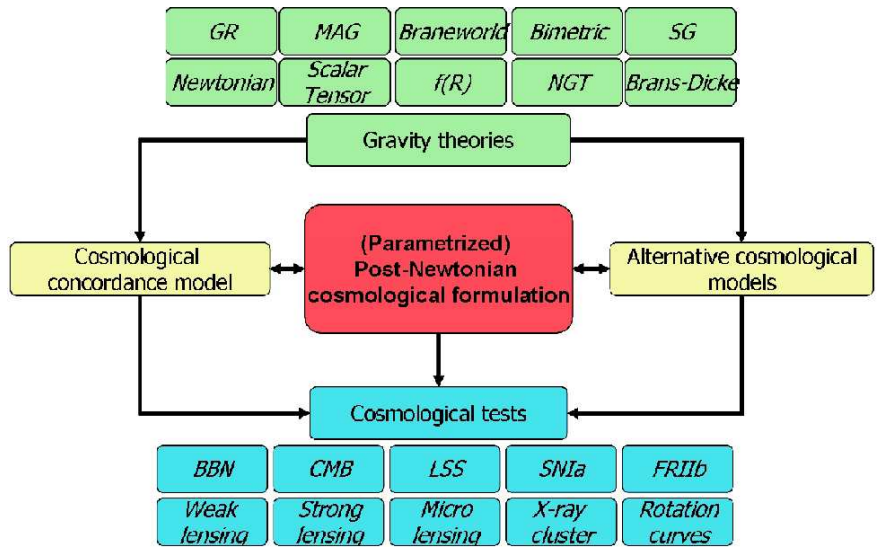}
  \caption{The great advantage of a systematic approximative framework is depicted here for a future parametrized post-Newtonian approximation in cosmology. In contrast to the current strategy of testing different gravitational theories by working out single cosmological tests one could quickly asses the viability of a given theory by comparing its predictions for some cosmological PPN parameters. As described in the text, we need a unified PM and PN inspired framework to cover all of the cosmological tests.}
  \label{models_fig}
\end{figure}

\section{Cosmology}

Let us now come to the situation in cosmology. As depicted in figure \ref{scales_fig} one currently applies different approximations on varying cosmological scales. From a more general standpoint the approximations used in cosmology can be traced back to the schemes described in the last section. The theory of cosmological perturbations, e.g., could also be viewed as a modified version of the post-Minkowskian approximation. The major difference is that one now starts from the cosmological Friedman-Lema\^{\i}tre-Robertson-Walker spacetime instead of the Minkowski spacetime. 

It is interesting that there has been made ample use of PM like approximation techniques in cosmology. In contrast an approximative scheme in the spirit of the previously described post-Newtonian approximation has not attracted much attention during the last years \cite{ShibataAsada1995,TakadaFutamase:1998}. This is even more surprising if one thinks of important cosmological tests like the formation of structure, in which the physical assumptions of weak gravitational fields and slow motions find a direct application. In the next section we briefly comment on the form of the equations of motion in the cosmological PN approximation. 

Finally, let us note that from figure \ref{scales_fig} it also becomes immediately clear that in cosmology we need a unification of both schemes, i.e.\ PM and PN, into one approximative framework which covers all scales.

\subsubsection{Post-Newtonian cosmology}

Recently we started the post-Newtonian part of the program in \cite{HwangNohPuetzfeld:2005,Puetzfeld:2005}. In order to provide a basis for future numerical structure formation simulations we derived the {\it cosmological} equations of hydrodynamics up to the first post-Newtonian order (1PN). They read\footnote{The dot ``$\cdot$'' denotes a partial derivative with respect to $t$.}:  

\begin{eqnarray}
0&=&c\left\{ \frac{1}{a^{3}}\left( a^{3}\rho \right) _{,t}+\left( \rho v^{a}\right) _{,a}\right\}   \nonumber \\
&&+\frac{1}{c}\left\{ \left( \rho \Pi \right) _{,t}+\dot{\rho}\left(a^{2}v^{2}+2U\right)\right.  \nonumber \\ 
&&\left. +3\frac{\dot{a}}{a}\left( p+2\rho U+\rho \Pi \right) +\rho \left[ a^{2}\left( v^{2}\right) _{,t}+3\dot{V}\right] \right.  \nonumber \\
&&\left. +\left[ \rho v^{a}\left( \Pi +a^{2}v^{2}\right) \right]_{,a}+\left( pv^{a}\right) _{,a}+\rho v^{a}\left( 3V-U\right) _{,a}\right.  \nonumber \\
&&\left.+2U\left(v^{a}\rho \right) _{,a}\right\} +O\left( c^{-3}\right),  \label{eom1} \\
0&=&\left\{ \left( \rho v^{a}v^{b}\right)_{,a}+\left( \rho v^{b}\right)_{,t}  \right.  \nonumber\\
&&\left. +\frac{1}{a^{2}}\left( p_{,b}-\rho U_{,b}\right)+5\frac{\dot{a}}{a}\rho v^{b}\right\}\nonumber\\
&&+\frac{1}{c^{2}}\left\{ \left( \rho \Pi v^{b}\right) _{,t}+\left( \rho \Pi v^{b}v^{a}\right) _{,a}+\left( pv^{b}\right) _{,t}\right.  \nonumber\\
&&+\left( pv^{b}v^{a}\right)_{,a}-V_{,b}\rho v^{2}+\rho a^{2}\left[ \left( v^{b}v^{a}v^{2}\right) _{,a} \right. \nonumber\\
&&\left. +\left(v^{b}v^{2}\right) _{,t}\right]+\frac{\rho }{a^{2}}U_{,b}\left[ 2\left( V-U\right) -\Pi -\frac{p}{\rho }-a^{2}v^{2}\right]  \nonumber\\
&&+2U\left[ \left( \rho v^{b}v^{a}\right)_{,a}+\left( v^{b}\rho \right)_{,t}\right]  \nonumber\\
&&+v^{b}\left[ \rho \left( 5\frac{\dot{a}}{a}\Pi +7\dot{a}av^{2}+10\frac{\dot{a}}{a}U+5V_{,a}v^{a}+U_{,a}v^{a} \right. \right. \nonumber \\
&&\left. \left. +\dot{U}+5\dot{V}\right)+ \dot{\rho}a^{2}v^{2}+5\frac{\dot{a}}{a}p+\rho _{,a}v^{a}a^{2}v^{2}\right]  \nonumber\\
&&\left. +\frac{1}{a^{2}}\left[ \frac{1}{2}\sigma _{,b}\rho -2Vp_{,b}+\rho \left( h_{0a,b}-h_{0b,a}\right) v^{a} \right. \right. \nonumber \\
&&\left. \left. -\rho \dot{h}_{0b}\right] \right\}+O\left( c^{-4}\right).  \label{eom2}
\end{eqnarray}

Here the quantities $U,V,\sigma, \upsilon,$ and $h_{0a}$ denote gravitational potentials from the different metric components. For details see \cite{Puetzfeld:2005}. As usual, the cosmological scale factor is denoted by $a$ and the quantities $\rho, \Pi, p, v^a$ represent the matter variables of the system, i.e.  energy-density, internal energy-density, pressure, and three velocity of the fluid elements. 

The equations (\ref{eom1}) and (\ref{eom2}) replace the continuity equation and Euler equations that are usually used in Newtonian studies. The 1PN equations should be used in future simulations to describe the first order corrections of General Relativity to cosmological hydrodynamics, for example in codes which study the formation of structure in the non-linear clustering regime. Although we expect that the post-Newtonian $c^{-2}$ corrections to the equations of motion are small in many situations, we cannot preclude that there are secular effects in the structure formation process. 

In other words, one still has to show, by performing a numerical simulation on the basis of the higher than Newtonian order equations, that the currently used Newtonian description is appropriate. 

The set of field equations to be solved in connection with the equations of motion (\ref{eom1}) and (\ref{eom2}) are given in \cite{Puetzfeld:2005}. The equations therein are kept in a form which closely resembles Chandrasekhar's approach \cite{Chandrasekhar:1965:1} in the non-cosmological case. In \cite{HwangNohPuetzfeld:2005} we also consider a mixture of post-Newtonian and perturbative methods and provide the final set of equations for several gauge conditions. At this point we would like to stress that, depending on the needs of the people who perform simulations, we would be glad to work out the optimal form of the field equations and equations of motion for numerical treatment.

\section{Problems \& Outlook}

Let us summarize that currently we do not have a quantitative estimate for the contribution of post-Newtonian effects to some very crucial cosmological tests. An important example being the formation of structures in the universe. Hence it is of utmost importance to study these effects numerically in a simulation in order to assess up to which level we can trust current simulations which only make use of the Newtonian limit. Up to 1PN order we derived all of the necessary equations and would be glad to assist other groups in the numerical implementation of these equations. 

On the theoretical side one of the open challenges of a cosmological PN approximation is linked to the need for conditions which clearly specify the realm of applicability of such an approximation. That means that we have to find a way to control the errors which we inevitably introduce when we treat a relativistic gravity theory in a Procrustean way. Another open issue in the cosmological case is the possible emergence of divergencies and thereby breakdown of the approximation at higher post-Newtonian orders. Such a behavior is already well known in the case of the PN approximation for isolated gravitating systems \cite{AndersonKegeles:1980}. Although it is unlikely that we need the improved accuracy of higher PN orders in any cosmological test in the near future, a thorough theoretical treatment should also address this fundamental issue.       

As depicted in figure \ref{models_fig}, the cosmological model building process would also greatly benefit from a parametrized post-Newtonian (PPN) framework. Such a scheme would allow for a rapid comparison of different theories by comparing their predictions for a set of standardized cosmological PPN parameters. To cover all cosmological tests one should aim for a framework which unifies PM and PN techniques. 
 
In view of a possible future PPN scheme for cosmology the question emerges whether General Relativity or, more precise, a metric theory of gravitation, is the appropriate foundation for such a scheme. As becomes clear from figure \ref{models_fig}, there is no lack of alternative theories which have been used to explain current cosmological observations. Some of these theories deviate significantly from GR. Therefore, our choice of the structure underlying a future PPN inspired cosmological framework should be guided by the need to incorporate as many richer theories as possible.

\begin{theacknowledgments}
I would like to thank the organizers of the conference, especially Hang Bae Kim (Hanyang University) for his generous support. I am also deeply indebted to Changbom Park (Korea Institute for Advanced Study), Jai-Chan Hwang (Kyungpook National University), and Hyerim Noh (Korea Astronomy and Space Science Institute) for making this trip to Korea possible and enjoyable in every aspect.   
\end{theacknowledgments}

\bibliographystyle{aipproc}

\begin{thebibliography}{28}
\expandafter\ifx\csname natexlab\endcsname\relax\def\natexlab#1{#1}\fi
\providecommand{\enquote}[1]{``#1''}
\expandafter\ifx\csname url\endcsname\relax
  \def\url#1{\texttt{#1}}\fi
\expandafter\ifx\csname urlprefix\endcsname\relax\def\urlprefix{URL }\fi
\providecommand{\eprint}[2][]{\url{#2}}

\bibitem[Einstein(1916)]{Einstein:1916:2}
A.~Einstein, \emph{Sitzungsb. der K\"{o}nig. Preuss. Akad} p. 688 (1916).

\bibitem[Lorentz and Droste(1917)]{LorentzDroste:1917}
H.~Lorentz, and J.~Droste, \emph{Versl. K. Akad. Wet. Amsterdam} \textbf{26},
  392 (1917).

\bibitem[Einstein et~al.(1938)]{EinsteinInfeldHoffman:1938}
A.~Einstein, L.~Infeld, and B.~Hoffmann, \emph{Ann. Math.} \textbf{39}, 65
  (1938).

\bibitem[Fock(1939)]{Fock:1939}
V.~Fock, \emph{Journal of Physics (USSR)} \textbf{1}, 81 (1939).

\bibitem[Einstein and Infeld(1940)]{EinsteinInfeld:1940}
A.~Einstein, and L.~Infeld, \emph{Ann. Math.} \textbf{41}, 455 (1940).

\bibitem[Einstein and Infeld(1949)]{EinsteinInfeld:1949}
A.~Einstein, and L.~Infeld, \emph{Can. J. Math.} \textbf{1}, 209 (1949).

\bibitem[Papapetrou(1951{\natexlab{a}})]{Papapetrou:1951}
A.~Papapetrou, \emph{Proc. Phys. Soc. Lond. A} \textbf{64}, 57
  (1951{\natexlab{a}}).

\bibitem[Papapetrou(1951{\natexlab{b}})]{Papapetrou:1951:2}
A.~Papapetrou, \emph{Proc. Phys. Soc. Lond. A} \textbf{64}, 302
  (1951{\natexlab{b}}).

\bibitem[Fock(1959)]{Fock:1959}
V.~Fock, \emph{The theory of space time and gravitation}, Pergamon Press, New
  York (Orig. 1955), 1959.

\bibitem[Soffel(1989)]{Soffel:1989}
M.~Soffel, \emph{Relativity in astronomy, celestial mechanics, and geodesy},
  Springer, Berlin, 1989.

\bibitem[Brumberg(1991)]{Brumberg:1991}
V.~Brumberg, \emph{Essential relativistic celestial mechanics}, Adam Hilger,
  Bristol, 1991.

\bibitem[Will(1993)]{Will:1993}
C.~Will, \emph{Theory and experiment in gravitational physics}, Cambridge
  University Press, Cambridge, 1993.

\bibitem[Kerr(1959{\natexlab{a}})]{Kerr:1959:1}
R.~Kerr, \emph{Nuovo Cim.} \textbf{13}, 469 (1959{\natexlab{a}}).

\bibitem[Kerr(1959{\natexlab{b}})]{Kerr:1959:2}
R.~Kerr, \emph{Nuovo Cim.} \textbf{13}, 492 (1959{\natexlab{b}}).

\bibitem[Kerr(1959{\natexlab{c}})]{Kerr:1959:3}
R.~Kerr, \emph{Nuovo Cim.} \textbf{16}, 673 (1959{\natexlab{c}}).

\bibitem[Anderson and Decanio(1975)]{AndersonDecanio:1975}
J.~Anderson, and T.~Decanio, \emph{1975} \textbf{6}, 197 (1975).

\bibitem[Anderson(1976)]{Anderson:1976:2}
J.~Anderson, \emph{Gen. Rel. Grav.} \textbf{7}, 643 (1976).

\bibitem[Ehlers(1977)]{Ehlers:1977}
J.~Ehlers, \emph{MPI-preprint MPI/PAE/Astro 138} p.~45 (1977).

\bibitem[Chandrasekhar(1965{\natexlab{a}})]{Chandrasekhar:1965:1}
S.~Chandrasekhar, \emph{Phys. Rev. Lett.} \textbf{14}, 241
  (1965{\natexlab{a}}).

\bibitem[Chandrasekhar(1965{\natexlab{b}})]{Chandrasekhar:1965:2}
S.~Chandrasekhar, \emph{Astrophys. J.} \textbf{142}, 1488 (1965{\natexlab{b}}).

\bibitem[Chandrasekhar(1969)]{Chandrasekhar:1969}
S.~Chandrasekhar, \emph{Astrophys. J.} \textbf{158}, 45 (1969).

\bibitem[Chandrasekhar and Nutku(1969)]{ChandrasekharNutku:1969}
S.~Chandrasekhar, and Y.~Nutku, \emph{Astrophys. J.} \textbf{158}, 55 (1969).

\bibitem[Chandrasekhar and Esposito(1970)]{ChandrasekharEsposito:1970}
S.~Chandrasekhar, and F.~Esposito, \emph{Astrophys. J.} \textbf{160}, 153
  (1970).

\bibitem[Shibata and Asada(1995)]{ShibataAsada1995}
M.~Shibata, and H.~Asada, \emph{Prog. Theor. Phys.} \textbf{94}, 11 (1995).

\bibitem[Takada and Futamase(1998)]{TakadaFutamase:1998}
M.~Takada, and T.~Futamase, \emph{Prog. Theor. Phys.} \textbf{100}, 315 (1998).

\bibitem[Hwang et~al.(2005)]{HwangNohPuetzfeld:2005}
J.~Hwang, H.~Noh, and D.~Puetzfeld, \emph{preprint}  (2005),
  \urlprefix\url{astro-ph/0507085}.

\bibitem[et~al(2005)]{Puetzfeld:2005}
D.~P. et~al, \emph{in preparation}  (2005).

\bibitem[Anderson and Kegeles(1980)]{AndersonKegeles:1980}
J.~Anderson, and L.~Kegeles, \emph{Gen. Rel. Grav.} \textbf{12}, 633 (1980).

\end{thebibliography}

\end{document}